\documentclass[aps]{revtex4}
\usepackage{color}
\usepackage{graphicx}
\usepackage{eurosym}
\usepackage{graphicx}
\usepackage{amsmath,amssymb}
\usepackage{color}
\usepackage[outdir=./]{epstopdf}
\usepackage{amsmath}
\usepackage{amsfonts}
\usepackage{amssymb}

\def\bc{\begin{center}}
\def\ec{\end{center}}

\def\beq{\begin{eqnarray}}
\def\eeq{\end{eqnarray}}

\def\bc{\begin{center}}
\def\ec{\end{center}}

\def\beq{\begin{eqnarray}}
\def\eeq{\end{eqnarray}}

\begin{document}

\title{Solar sail with superconducting circular current-carrying wire}
\author{ V. Ya. Kezerashvili$^{1}$ and R. Ya. Kezerashvili$^{1,2}$}
\affiliation{\mbox{$^{1}$Physics Department, New York City College
of Technology, The City University of New York,} \\
Brooklyn, NY 11201, USA \\
\mbox{$^{2}$The Graduate School and University Center, The
City University of New York, \\
New York, NY 10016, USA }\\
}
\date{\today }

\begin{abstract}
\textbf{Background:} A solar sail presents a large sheet of low areal
density membrane and is the most elegant propellant-less propulsion system
for the future exploration of the Solar System and beyond. By today the
study on sail membrane deployment strategies has attracted considerable
attention.

\textbf{Goal:} In this work we present an idea of the deployment and
stretching of the circular solar sail. We consider the superconducting
current loop attached to the thin membrane
and predict that a superconducting current loop can deploy and stretch the
circular solar sail membrane.

\textbf{Method:} In the framework of a strict mathematical approach based on
the classical electrodynamics and theory of elasticity the magnetic field
induced by the superconducting current loop and elastic properties of a
circular solar sail membrane and wire loop are analyzed. The formulas for
the wire and sail membrane stresses and strains caused by the current in the
superconducting wire are derived.

\textbf{Results:} The obtained analytical expressions can be applied to a
wide range of solar sail sizes. Numerical calculations for the sail of
radius of 5 m to 150 m made of CP1 membrane of the
thickness of 3.5 $\mu m$ attached to Bi$-$2212 superconducting wire with the
cross-section radius of 0.5 mm to 10 mm are presented. Calculations are performed for the
engineering current densities of 100 A/mm$^{2}$ to 1000 A/mm$^{2}$.

\textbf{Conclusion:} Our calculations demonstrate the feasibility of the
proposed idea for the solar sail deployment for the future exploration of
the deep space by means of the light pressure propellent.
\end{abstract}

\maketitle

\section{Introduction}

Today the \emph{de facto }chemical propulsion rocket remains the main space
exploration vehicle. However, this propulsion system
is faced with several difficulties such as: i. the necessity to transport
fuel on a board imposes prohibitive requirements on mass/payload ratio and
economic coast; ii. the maximum speed that a rocket can reach is limited by
the rocket equation. The question arises: Can we use natural astrophysical
sources as a propulsion mechanism for space exploration? There are a variety
of suggestions alternative to the traditional propulsion system that must
carry fuel on the vehicle: i. a solar sail that is accelerated by the Sun
electromagnetic flux; ii. magnetic sail \cite{Zubrin1991,Zubrin1993}; iii.
electric sail \cite{Janhunen2004,Janhunen2007}. The magnetic and electric
sails deflect and extract the momentum from the solar wind particles
using the induced on a board magnetic and electric fields, respectively. A
spacecraft based on such a propulsion mechanism no longer need to carry the
mass of a propulsion system and would not require refueling missions to
increase the longevity of the fuel-propelled spacecraft.

A solar sail is the most elegant propellant-less propulsion system for the
future exploration of the Solar System and beyond. A solar sail is a large
sheet of low areal density material that captures and reflects the Sun
electromagnetic flux as a means of acceleration. Let's give a short overview
of the concept of solar sailing. Over 150 years ago, in 1873 James Clerk
Maxwell \cite{Maxwell} in his famous \textquotedblleft A Treatise on
Electricity and Magnetism\textquotedblright\ published by the Oxford
University Press, theoretically predicted that electromagnetic radiation
exerts pressure upon any surface exposed to it. It took 27 years before
Lebedev \cite{Lebedev} and independently Nicholas and Hill \cite%
{Nichols1901,Nichols1903} presented the first experimental evidence that
confirmed Maxwell's prediction that light had a measurable pressure in
agreement with Maxwell's equations.

Interesting enough, in 1915 following the experimental verification of the
solar radiation pressure, Yakov Perelman, a Soviet science writer and author
of many popular science books, in his book titled "Interplanetary Journeys"
\cite{Perelman1915} proposed that solar radiation pressure could be used for
the propulsion of solar sail spacecraft. However, because the solar
radiation pressure is too small, he concluded that such a spacecraft should
be mostly unrealistic.
The concept of solar sailing was articulated as an engineering principle in
the early 1920s by Konstantin Tsiolkowsky along with Fridrikh Tsander. In
1924 Tsander \cite{Tsander1924} promoted Tsiolkovsky's work and developing
it further. Although the basic idea behind solar sailing appears simple,
challenging engineering problems must be solved.

After the launch of the first artificial satellite Sputnik in 1957 the
question of the importance of the effect of solar radiation on the orbital
motion of satellites was raised. In general, the perturbing effects of solar
radiation pressure on satellite orbits have been considered by celestial
mechanicians to be negligible. However, the studies of Parkinson et al. \cite%
{Parkinson1960} and Musen \cite{Musen1960} pointed to the importance of the
effects of solar radiation pressure on Earth satellite orbits. The Vanguard
I was an American satellite that was the fourth artificial satellite
launched into Earth orbit. The difference between observed and theoretical
values of perigee height for the Vanguard I satellite has suggested a
reexamination of radiation pressure as a possible source of the discrepancy.
An investigation of the effect of solar radiation pressure on the motion of
an artificial satellite was reported in Ref. \cite{Musen21960}. The theory
has been applied to the orbit of the Vanguard I satellite for which the
discrepancies between the theoretical and experimental orbits were observed.
The inclusion of the effect of radiation pressure led to close agreement
between the orbit data and the theoretical results for Vanguard I. It was
also found that for representative values of the orbit elements of the Echo
I balloon satellite, solar radiation can in fact produce orbit perturbations
of the order of hundreds of kilometers in a few months. The successful
launch of the Echo I balloon on August 12, 1960 \cite{Pezditz1962} provided
the first definitive test of the effect of solar radiation pressure on the
satellite orbits. During the first 12 days the motion of the Echo
communications satellite clearly confirmed predictions of the influence of
solar radiation pressure. During this time, solar pressure reduced perigee
height by 44 km \cite{Shapiro1960}. Calculations show that, at a mean
altitude of 1600 km, radiation pressure can displace the orbit of the
30.5-meter diameter Echo balloon satellite at rates up to 6 km per day, the
orbit of the inflatable 3.66-meter Beacon satellite at 1.1 km per day. For\
the certain resonant conditions, this effect accumulates and drastically
affects the satellite's lifetime.

Only 110 years after the experimental measurements of the solar radiation
pressure a JAXA team reported the injection of the world's first
interplanetary solar sail, Japan's $\sim $ 200 m$^{2}$ IKAROS, which
demonstrated the feasibility of spacecraft propulsion by solar radiation
pressure \cite{Mori2011,ASRKez2011}.

The propulsion using a solar sail has three primary and complementary foci: i.
finding low areal density material that allows the deployment of the sail
close to the Sun to utilize the maximum possible acceleration due to the
solar radiation pressure; ii. the area of the solar sail made of a low areal
density material needs to be maximized to increase the solar thrust; iii. the development of the mechanism for the deployment and stretching the large size of solar sail membrane. 

The study on sail membrane deployment strategies has attracted considerable
attention. An important question that arises in the context of deployable
solar sail structures is their weight and stability. Many different systems
have been previously considered for the sail opening. Each system was
characterized by the presence of guide rollers, electromechanical actuation
devices, or composite booms \cite{9}. The deployment is usually performed by
uniaxial mechanisms, such as a telescopic boom, the extendable masts, the
deployable booms, the inflatable booms, the centrifugal force that renders a
spin-type deployment mechanism (See Refs. \cite%
{Deployment4,Deployment1,Deployment2,Deployment3} and references therein).
We cite these works, but the recent literature on the subject is not limited
by them. Recently an alternative method for the solar sail self-deployment
based on shape memory alloys was suggested \cite{MemorySail, MemorySail2},
where the authors use shape memory alloys as mechanical actuators for solar
sail self-deployment instead of heavy and bulky mechanical booms. Most
recently a torus-shaped sail consisting of a reflective membrane attached to
an inflatable torus-shaped rim was suggested \cite{KezASR2021}. The sail
deployment from its stowed configuration is initiated by the introduction of
the inflation pressure into the toroidal rim. However, in the actual
deployment technology of the solar sail, the main limit is still the high
weight of the system and the complexity of the deployment mechanism for the
solar sail surface.

We propose a circular superconducting current-carrying wire attached to a
circular solar sail to achieve the solar sail deployment and stretching. To the
best of our knowledge, such a configuration was not considered so far,
although the magnetic and elastic properties of a circular current-currying
wire alone constitute a well-known problem of the classical electrodynamics
\cite{Maxwell,Jeans27,Sommerfeld48,
Abraham32,Panofsky62,Landau8,Smythe89,Greiner98,Jackson98} and the theory of
elasticity \cite{Timoshenko51,Landau7,Boyko94}.

This article is organized in the following way. In Sec. \ref{theory} within
the framework of the classical electrodynamics, we consider the magnetic
field of the thin superconducting wire which generates magnetic self-forces
that lead to the deployment of an ultra-lightweight circular sail membrane
attached to the wire. The stress and strain of the circular membrane under
the uniformly distributed force applied to the membrane edge as well as the
stress and strain in the wire-membrane combination are considered within the
theory of elasticity. Results of calculations and discussion are presented
in Sec. \ref{results}. The concluding remarks follow in Sec. \ref%
{conclusions}.

\section{Circular Current Wire Attached to Circular Membrane}

\label{theory}

\begin{figure}[b]
\centering
\includegraphics[width=10cm]{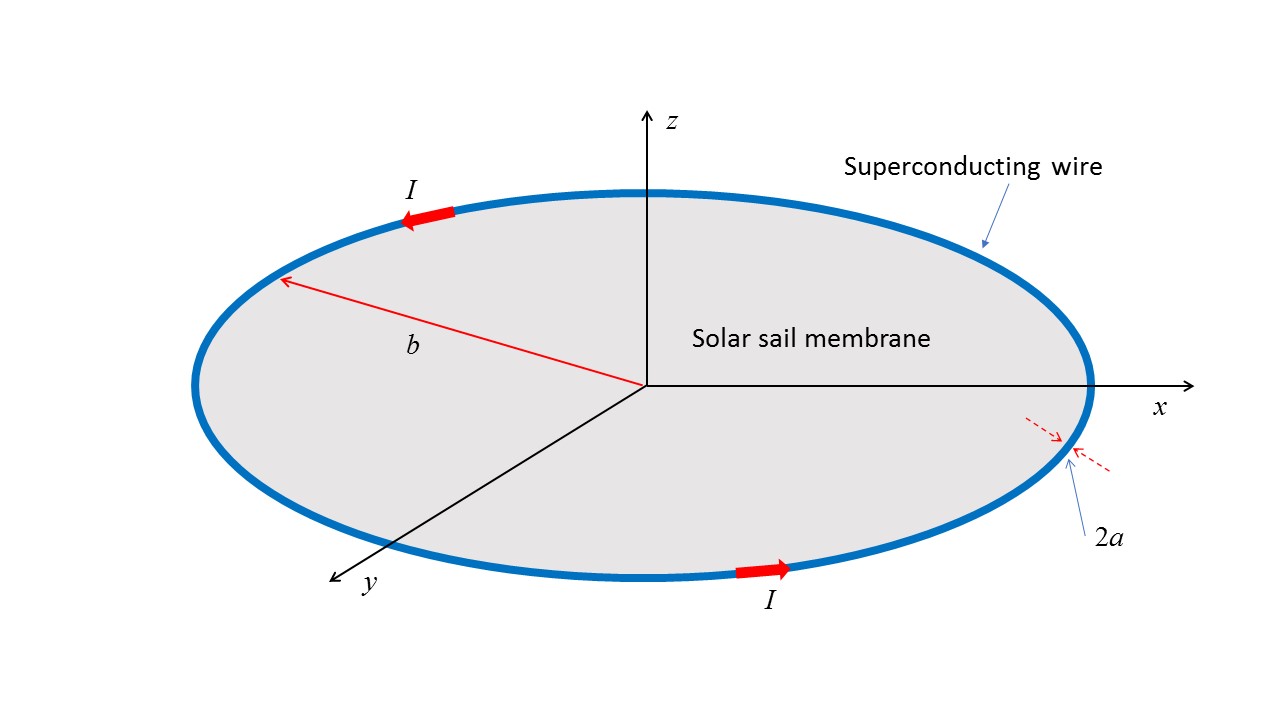}
\caption{(Color online) A schematic of the circular solar sail of a radius $%
b $ attached to a superconducting circular wire of the cross-section radius $%
a$ currying steady-state current $I$ and laying in $x-y$ plane. The solar
sail is stretched by self-forces generated by the magnetic field induced by
the current-carrying wire on itself. The radius $b$ is significantly bigger
than the cross-section radius of the wire $a$: $b\gg a$. The figure is not
to scale.}
\label{Fig1a}
\end{figure}

A schematic of the circular solar sail of a radius $b$ attached to a
superconducting circular wire of a circular cross-section radius $a$
currying steady-state current $I$ is presented in Fig. \ref{Fig1a}. The
radius of the circular wire $b$ is significantly greater than the radius of
the wire cross-section $a$: $b\gg a$. The solar sail membrane is stretched
by self-forces generated by the magnetic field induced by the
current-carrying wire on itself. In this Section we consider the circular
current-carrying wire and the magnetic self-forces induced by the current.
After that, the detailed consideration of the stress and strain in the
circular membrane resulting from uniformly distributed force applied to the
membrane edge is presented. Finally, the combined system of the circular
membrane with the attached superconducting wire is considered.

\subsection{\protect\bigskip Circular Current Wire}

\label{A}

\begin{figure}[b]
\centering
\includegraphics[width=7cm]{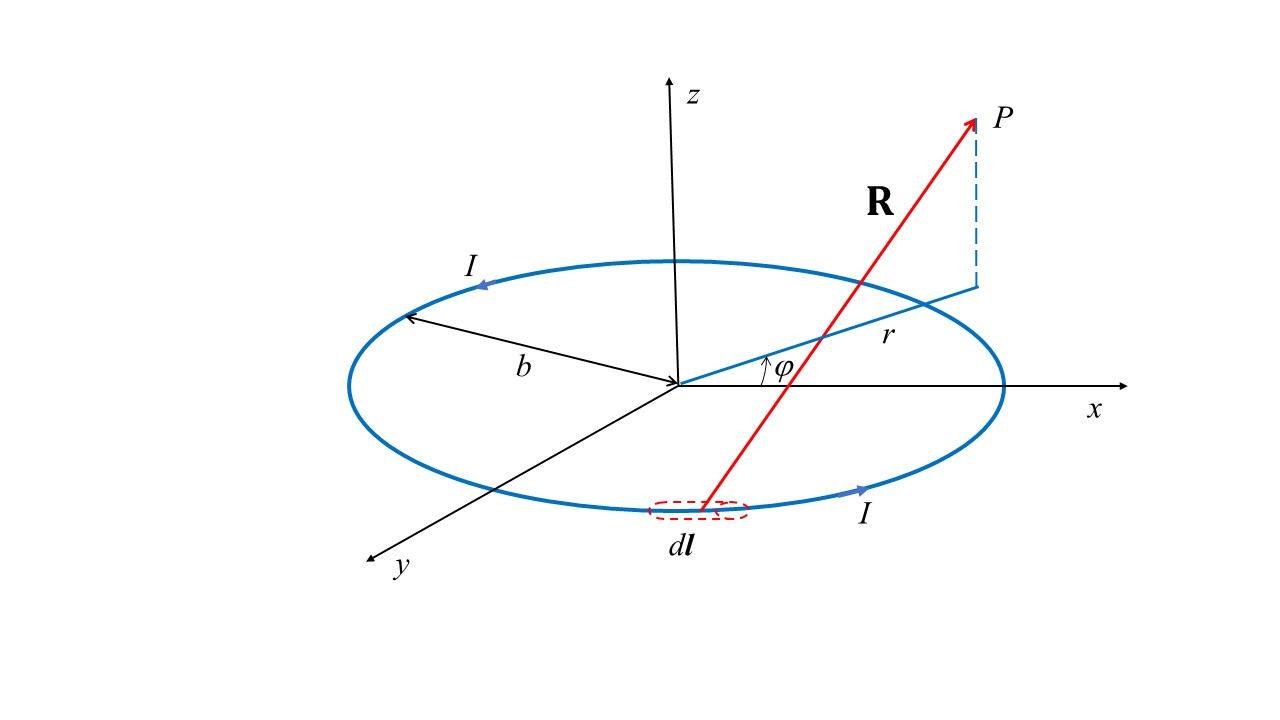} \includegraphics[width=7cm]{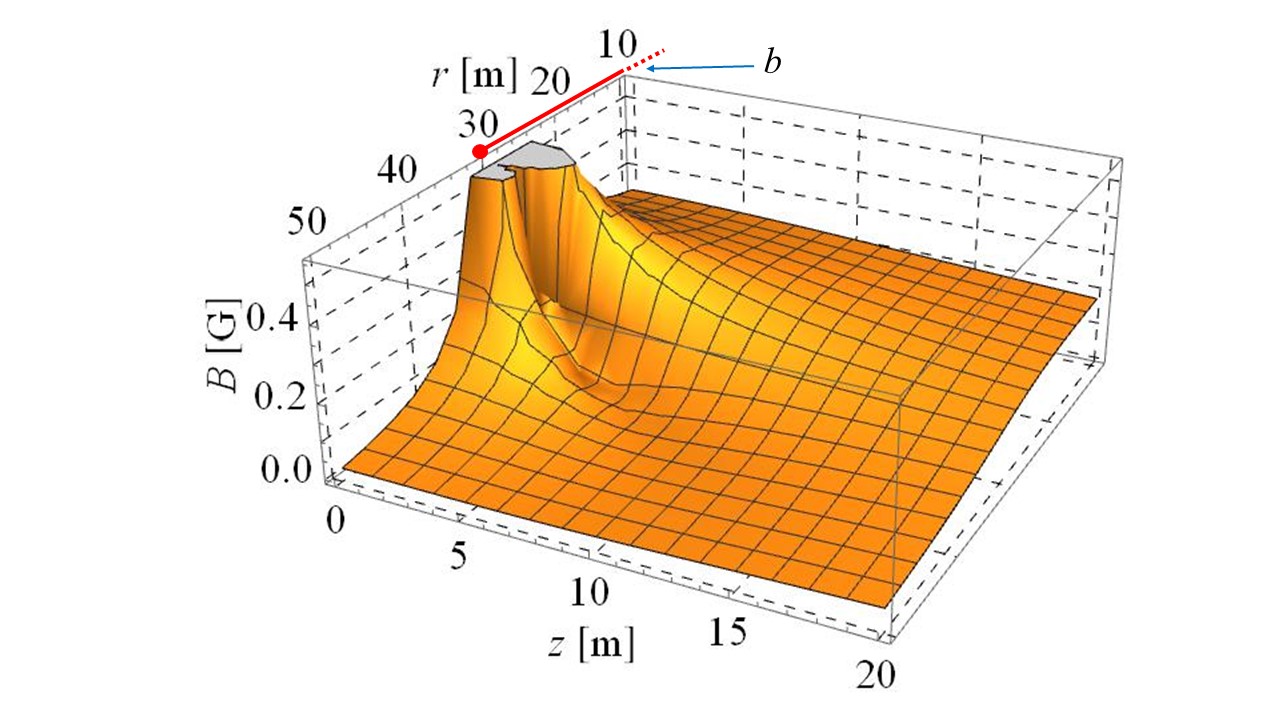}
\caption{(Color online) Left panel: The circular wire laying in $x-y$ plane,
centered at the origin and currying steady-state current $I$. The radius of
the wire is $b$. $P$ indicates any point in the surrounding space where the
magnetic field is induced by the wire current. Right panel: The magnetic
field of the circular current-carrying wire as a function of $r$ and $z$.
The logarithmic singularity of the magnetic field at $r=b$ is indicated by
the point. The figure is not to scale: $b\gg a$.}
\label{Fig1}
\end{figure}

Let us consider a planar circular wire laying in $x-y$ plane, centered at
the origin and currying steady-state current $I$\ as shown in Fig. \ref{Fig1}%
. We assume that the permeability $\mu =\mu _{0}$, where $\mu _{0}=4\pi
\times 10^{-7}$T$\cdot $m$/$A is the permeability of the free space, in the
whole space including the volume of the wire, means that both the wire and
the surrounding medium are nonpermeable. If the wire is sufficiently thin
and the magnetic field of interest is only in surrounding space the
thickness of the wire can be neglected \cite{Landau8}. One can say that such
a wire carries a \textit{linear current}.
Let us answer the question: can one find the force acting on the linear
current due to its own magnetic field? For a system of any linear currents
the vector potential $\mathbf{A}$ and magnetic induction $\mathbf{B}$ in the
surrounding space become \cite{Jackson98, Landau8} :

\begin{align}
\mathbf{A}& \mathbf{=}\frac{\mu _{0}I}{4\pi }\oint \frac{d\mathbf{l}}{R},
\label{A potential} \\
\mathbf{B}& \mathbf{=}\frac{\mu _{0}I}{4\pi }\oint \frac{d\mathbf{l\times R}%
}{R^{3}},  \label{B induction}
\end{align}%
where for the vector potential uniqueness the Coulomb's gauge is assumed: $%
div$ $\mathbf{A}=0$. In Eqs. (\ref{A potential}) and (\ref{B induction}) $I$
is the total current in the wire and $\mathbf{R}$ shown in Fig. \ref{Fig1}
is the radius-vector from the current element $d\mathbf{l}$ to the point $P$%
, where $\mathbf{A}$ and $\mathbf{B}$ are observed.
By introducing the cylindrical coordinates $r$, $\varphi $, $z$ (Fig. \ref%
{Fig1}), due to axially symmetric 
current the vector Eq. (\ref{A potential}) can be reduced to single scalar
one:

\begin{equation}
A_{r}=0,\text{ }A_{z}=0,\text{ }A_{\varphi }=A(r,z).  \label{A component}
\end{equation}%
For Eq. (\ref{B induction}) 
we have

\begin{eqnarray}
B_{r} &=&-\frac{1}{r}\frac{\partial }{\partial z}\left( rA_{\varphi }\right)
+\frac{1}{r}\frac{\partial }{\partial \varphi }\left( A_{z}\right) =-\frac{%
\partial A_{\varphi }}{\partial z},\text{ }  \notag \\
B_{z} &=&-\frac{1}{r}\frac{\partial }{\partial \varphi }\left( A_{r}\right) +%
\frac{1}{r}\frac{\partial }{\partial r}\left( rA_{\varphi }\right) =\frac{1}{%
r}\frac{\partial }{\partial r}(rA_{\varphi }),\text{ }  \notag \\
B_{\varphi } &=&\frac{\partial }{\partial z}\left( A_{r}\right) -\frac{%
\partial }{\partial r}\left( A_{z}\right) =0.  \label{B components}
\end{eqnarray}%
It is worth noting that since we neglect the thickness of the wire, no
boundary conditions at its surface need to be applied.
The exact analytical solutions of Eqs. (\ref{A component}) and (\ref{B
components}) are well known \cite{Landau8, Smythe89}
\begin{align}
A_{\varphi }& =\frac{\mu _{0}I}{\pi k}\left( \frac{b}{r}\right) ^{1/2}\left[
\left( 1-\frac{1}{2}k^{2}\right) K(k)-E(k)\right] ,  \label{A_f} \\
B_{r}& =\frac{\mu _{0}I}{2\pi }\frac{z}{r\left[ \left( b+r\right) ^{2}+z^{2}%
\right] ^{1/2}}\left[ -K(k)+\frac{b^{2}+r^{2}+z^{2}}{\left( b-r\right)
^{2}+z^{2}}E(k)\right] ,  \label{B_r} \\
B_{z}& =\frac{\mu _{0}I}{2\pi }\frac{1}{\left[ \left( b+r\right) ^{2}+z^{2}%
\right] ^{1/2}}\left[ K(k)+\frac{b^{2}-r^{2}-z^{2}}{\left( b-r\right)
^{2}+z^{2}}E(k)\right] .  \label{B_z}
\end{align}%
In Eqs. (\ref{A_f}) - (\ref{B_z})
\begin{equation}
K(k)=\int\limits_{0}^{\pi /2}\frac{d\theta }{\sqrt{1-k^{2}\sin ^{2}\theta }}%
\quad \text{and }\quad E(k)=\int\limits_{0}^{\pi /2}\sqrt{1-k^{2}\sin
^{2}\theta }d\theta ,
\end{equation}%
%
%
are complete elliptical integrals of the first and second kind, respectively
\cite{Abramowitz, Ryzhik}. The argument $k$ of the elliptic integrals $K(k)$
and $E(k)$ is defined through $k^{2}=\frac{4br}{\left( b+r\right) ^{2}+z^{2}}
$ and $\theta =\frac{1}{2}\left( \varphi -\pi \right) .$

It is easy to obtain well-known results given in the general physics text
books for the magnetic field along $z$ axes and at the center of the
circular loop. At $z$ axes ($r=0$) $B_{r}=\underset{r\rightarrow 0}{\lim }%
B_{r}=0$ \cite{Smythe89,Abramowitz} and $B_{z}=\frac{\mu _{0}b^{2}I}{2\left(
b^{2}+z^{2}\right) ^{3/2}}$, while at the center of the circular loop $%
B_{r}=0$ and $B_{z}=\frac{\mu _{0}I}{2b}$. However, at the wire when $r=b$
and $z=0$, $k=1$ and, therefore, $K=\int\limits_{0}^{\pi /2}\frac{d\theta }{%
\cos \theta }=\ln \left\vert \tan \frac{\pi }{2}\right\vert $ diverges
logarithmically, so is the magnetic field. The right panel in Fig. \ref{Fig1}
shows an example of the distribution of magnetic field as a function of the $%
r$ and the distance $z$ from the $x-y$ plane. The calculations are performed
for $I=1000$ A, $b=30$ m. Along the circumference of the radius $r=b$ when $%
z=0$ the magnetic field becomes singular. The force being a product of the
current, circumference, and magnetic field becomes singular also. This
result means that an attempt to calculate the force acting on the linear
current due to its own magnetic field leads to the logarithmic divergence.
Thus, to find the magnetic self-force on the circular wire due to its own
magnetic field the cross-section of the current-caring wire can not be
ignored.

\subsection{Current-carrying wire self-forces}

\label{B}

The total energy of the magnetic field of a single wire carrying a constant
current $I$ is \cite{Landau8, Smythe89} :
\begin{equation}
U=\frac{1}{2}LI^{2},  \label{Total E}
\end{equation}%
where $L$ is the self-inductance of the wire. Again let us assume that the
wire and surrounding media are nonpermeable ( $\mu =\mu _{0}$). To establish
a current $I$ in the wire the change of energy of sources of electromotive
force (\textit{EMF}) are $-LI^{2}$ \cite{Abraham32}, so that the total free
energy of the system of the \textit{EMF }and wire is
\begin{equation}
\widetilde{U}=-\frac{1}{2}LI^{2}.
\end{equation}%
Thus, the total magnetic energy of the wire is
\begin{equation}
U=\frac{1}{2}LI^{2}=-\widetilde{U}.
\end{equation}%
The forces acting on the wire are \cite{Landau8,Jackson98}
\begin{equation}
F_{q}=\left( -\frac{\partial \widetilde{U}}{\partial q}\right) _{I}=\left(
\frac{\partial U}{\partial q}\right) _{I}=\frac{1}{2}I^{2}\left( \frac{%
\partial L}{\partial q}\right) _{I},  \label{Forces}
\end{equation}%
where $q$ is a generalized displacement of the wire and the partial
derivative is taken under fixed current. 
These forces act as follows. For the fixed $I$ the total free energy $%
\widetilde{U}$ reaches the minimum. The latter means that the forces acting
on the wire will tend to increase wire self-inductance. $L$ having the
dimension of the length times $\mu _{0}$, is proportional to the size of the
wire. Therefore, the size of the circular wire increases under the action of
the magnetic self-field \cite{Landau8}.

The self-inductance of a wire can not be calculated using Newmann's formula
for the mutual inductance between $i$ and $k$ current-carrying wires \cite%
{Panofsky62}
\begin{equation}
L_{ik}=\frac{\mu _{0}}{4\pi }\oint \oint \frac{d\mathbf{l}_{i}d\mathbf{l}_{k}%
}{r_{ik}}
\end{equation}%
due to the logarithmic divergence at $r_{ik}\rightarrow 0$ arising from the
fact that the integrals have to be taken over the same contour. The
self-inductance is much more difficult to calculate. The standard way \cite%
{Landau8,Smythe89,Jackson98} is to calculate the energy of the magnetic
field over the whole nonpermeable space $V$ as
\begin{equation}
U=\frac{1}{2\mu _{0}}\int B^{2}dV
\end{equation}%
and then obtain $L$ from (\ref{Total E}) as $L=\frac{2U}{I^{2}}$. The
self-inductance of a current loop of the circular cross-section of radius $a$%
, length $C$ and a projected area $A$ (loop can be none planar) is \cite%
{Landau8,Jackson98}:

\begin{equation}
L\approx \frac{\mu _{0}}{4\pi }C\left( \ln \frac{\varsigma A}{a^{2}}+\frac{1%
}{2}\right) ,  \label{LLandau}
\end{equation}%
where $a\ll \frac{C}{2\pi }$\ or $a\ll A^{1/2}$and $\varsigma $ is a
unitless constant of the order of 1. Equation (\ref{LLandau}) is
logarithmically accurate and leads to an important conclusion: \textit{in
equilibrium a flexible closed current loop of any shape will tend to take
the shape spanning the maximum area }$A$\textit{\ for the fixed length of
the wire }$C$\textit{, which is the area of the circle }$A=\frac{C^{2}}{4\pi
}$. In other words the equilibrium shape of the flexible current-carrying
wire is a perfect circle due to the magnetic force self-action. It is worth
mentioning that due to inaccurate folding of the flexible wire it can be
bent and have kinks and as a result, any excess rigidity in any part of the
wire, will prevent wire to take the perfect circle shape. This does not
contradict the conclusion above, because in the general case the free energy
includes non-magnetic parts and may take a minimum with kinks remaining.

The self-inductance of the current-carrying circular wire of radius $b$ with
the radius $a$ of the cross-section area when $a\ll b$ is \cite%
{Abraham32,Landau8,Smythe89,Jackson98}:
\begin{equation}
L=\mu _{0}b\left( \ln \frac{8b}{a}-\frac{7}{4}\right) .  \label{LSC}
\end{equation}
Equation (\ref{LSC}) includes the contributions to the self-inductance due
to the field inside the wire $L_{i}$ and the field outside the wire $L_{0}$.
For type I superconductors and type II superconductors with the magnetic
field below the first critical field, the field is expelled from the volume
of the wire ($L_{i}=0$), so (\ref{LSC}) as shown in Refs. \cite{Landau8,Fok}
becomes
\begin{equation}
L=\mu _{0}b\left( \ln \frac{8b}{a}-2\right) .
\end{equation}%
For type II superconducting wire above the first critical field, the
magnetic field penetrates the wire into a regular array of vortices \cite%
{Abrikosov} and $L$ is still expressed by Eq. (\ref{LSC}). The same is true
for modern superconducting wires consisting of the regular array of
individual superconducting filaments bundeled together \cite{CSWire}.
\begin{figure}[b]
\centering
\includegraphics[width=5.0cm]{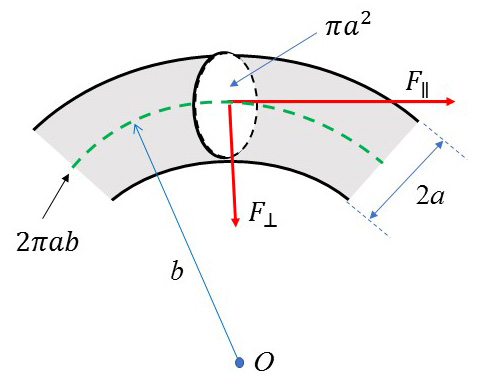}
\caption{(Color online) A schematic for the action of forces $F_{\parallel }$
and $F_{\perp }$ on the circular wire element. The force $F_{\parallel }$
acts on the cross-section area $\protect\pi a^{2}$ along the axis of the
wire. For visibility, the cross-section of the wire is given in 3D format.
The force $F_{\perp }$ acts on the area $2\protect\pi ab$ perpendicular to
the plane of the figure (shown by the dashed curve) along the normal to the
axis. The figure is not to scale: $b\gg a$. }
\label{Fig3}
\end{figure}

Now let us consider the self-action effect of the magnetic field on the
wire. The generalized displacements of the wire are the circumference $2\pi
b $ and the radius of the wire cross-section $a$. The variation of the
length of the circular loop, $2\pi b$, is responsible for the force $%
F_{\parallel }$ acting along the axis of the wire as shown in Fig. \ref{Fig3}%
, causing the tensile stress $\sigma _{\parallel }$ along it. The variation
of the radius of the wire leads to the force\ $F_{\perp }$ normal to the
axis and causing the stress $\sigma _{\perp }$ compressing the wire \cite%
{Landau8}. Following Ref. \cite{Landau8} and using Eqs. (\ref{Forces}) and (%
\ref{LSC}) we obtain

\begin{eqnarray}
F_{\parallel } &=&\frac{1}{2}I^{2}\frac{\partial L}{\partial \left( 2\pi
b\right) }=\frac{\mu _{0}}{4\pi }I^{2}\left( \ln \frac{8b}{a}-\frac{3}{4}%
\right) =\pi a^{2}\sigma _{\parallel }, \\
F_{\perp } &=&\frac{1}{2}I^{2}\frac{\partial L}{\partial a}=-\frac{1}{2}%
\frac{\mu _{0}b}{a}I^{2}=2\pi ab\sigma _{\perp },
\end{eqnarray}%
where
\begin{eqnarray}
\sigma _{\parallel } &=&\frac{\mu _{0}}{4\pi ^{2}}\frac{I^{2}}{a^{2}}\left(
\ln \frac{8b}{a}-\frac{3}{4}\right) ,  \label{SigmaP} \\
\sigma _{\perp } &=&-\frac{\mu _{0}}{2\pi }\frac{I^{2}}{a^{2}}.
\label{Sigmap}
\end{eqnarray}

The stresses $\sigma _{\parallel }$\ \ and $\sigma _{\perp }$ will strain
the wire leading to the relative elongation of the wire $\Delta b_{w}/b$,
which can be obtained in the framework of the classic theory of elasticity.
Following Refs. \cite{Landau8,Timoshenko51,Landau7} one can obtain
\begin{equation}
\frac{\Delta b_{w}}{b}=\frac{1}{E_{w}}(\sigma _{\parallel }-2\nu _{w}\sigma
_{\perp }),  \label{delB}
\end{equation}%
where $E_{w}$ and $\nu _{w}$ are the Young modulus of elasticity and Poisson
ratio of the wire, respectively. In accordance to (\ref{SigmaP}) and (\ref%
{Sigmap}), Eq. (\ref{delB}) gives the change of the wire loop's radius $b$:

\begin{equation}
\Delta b_{w}=\frac{\mu _{0}}{4\pi ^{2}}\frac{I^{2}}{a^{2}}\frac{b}{E_{w}}%
\left( \ln \frac{8b}{a}-\frac{3}{4}+2\pi \nu _{w}\right) .  \label{DelB}
\end{equation}%
\begin{figure}[t]
\centering
\includegraphics[width=14.0cm]{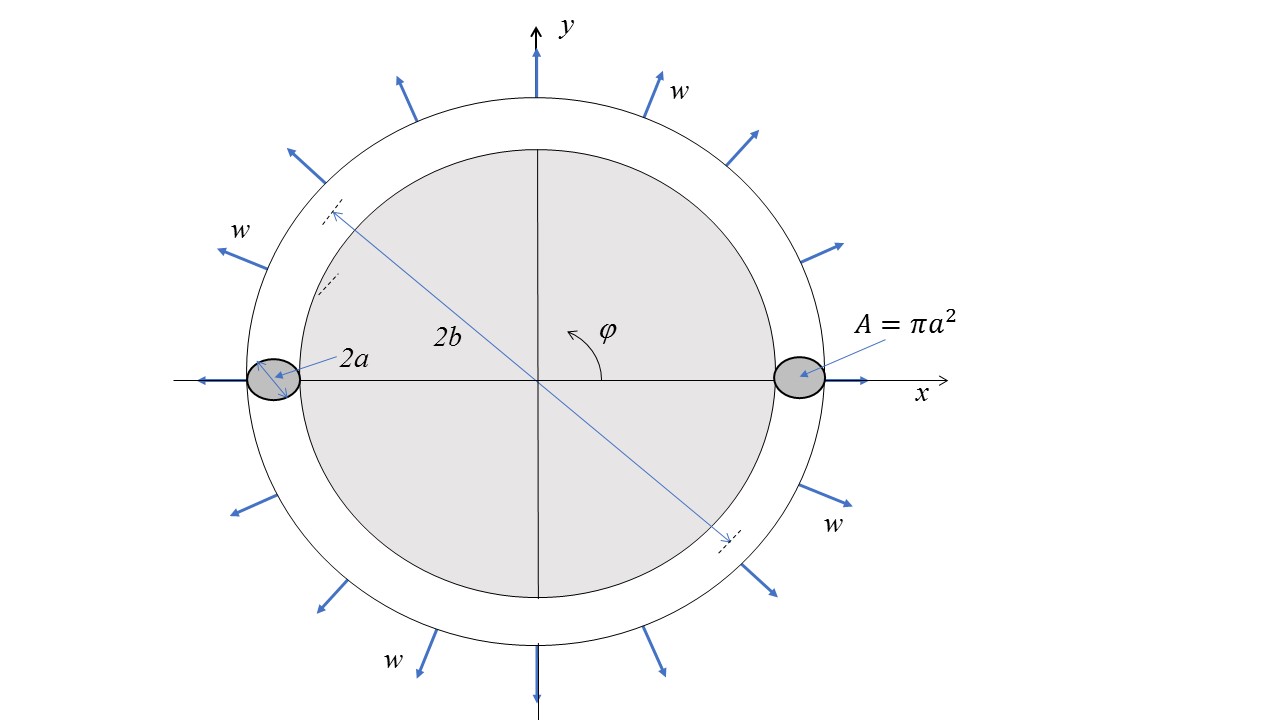}
\caption{(Color online) A schematic for the action of uniform radial force $%
\protect\omega $ acting per unit of circumferential length on the circular
current-carrying wire. For visibility, the cross-section of the wire is
given in 3D format. The figure is not to scale: $b\ll a$. }
\label{Fig4a}
\end{figure}
Let's consider the task of finding a radial force uniformly distributed
along the wire circumference which causes the change of the wire radius
equal to (\ref{DelB}). The wire under uniform radial force $w$ per unit of
circumferential length is shown in Fig. \ref{Fig4a}. The final result of
this classical problem is presented in Ref. \cite{Young} Table 9.2, case 12
and reads:

\begin{equation}
\Delta b_{w}=-\frac{wb^{4}}{2\pi E_{w}\mathfrak{I}}\left[ k_{1}\left( \frac{%
\pi ^{2}\sin \theta }{2}-\sin \theta +\theta \cos \theta \right) +k_{2}\pi
\left( \theta -\sin \theta -2\right) +2k_{2}^{2}\left( \pi -\theta -\sin
\theta \right) \right] .  \label{DbYoung}
\end{equation}%
In Eq. (\ref{DbYoung}) $\mathfrak{I}=\frac{\pi a^{4}}{4}$ is the moment of
inertia of the wire cross-section. In our case of the force distributed
uniformly throughout the wire circumference the angle $\theta $ equals 0. In
the case of $\theta =0$ the term with $k_{1}$ in the square brackets of Eq. (%
\ref{DbYoung}) is vanishing and only the terms with $k_{2}=1-\alpha $, where
$\alpha =\frac{\mathfrak{I}}{\pi a^{2}b^{2}}$ is the hoop-stress deformation
factor, will survive. Substitution of the moment of inertia of the wire
cross-section $\mathfrak{I}$ in the latter expression gives $\alpha =\frac{1%
}{4}\left( \frac{a}{b}\right) ^{2}\ll 1$. Therefore, within $\left( \frac{a}{%
b}\right) ^{2}$ accuracy Eq. (\ref{DbYoung}) reads:

\begin{equation}
\Delta b_{w}=\frac{wb^{2}}{\pi E_{w}a^{2}}.  \label{Dbfinal}
\end{equation}%
The same result for $\Delta b_{w}$ can be obtained with much less efforts
following Ref. \cite{Timoshenko51} as it is illustrated in Fig. \ref{Fig4a}
by calculating tensile stress $\sigma _{t}$ in the wire cross-section due to
the uniformly distributed radial force $w$
\begin{equation}
\sigma _{t}=\frac{1}{2\pi a^{2}}\int\limits_{0}^{\pi }wb\sin \varphi
d\varphi =\frac{wb}{\pi a^{2}}
\end{equation}%
and equating it to the strain $\frac{\Delta b_{w}}{b}$ by considering the
linear response according to the Hook's law
\begin{equation}
\sigma _{t}=E_{w}\frac{\Delta b_{w}}{b}\equiv \frac{wb}{\pi a^{2}}.
\label{26}
\end{equation}%
The latter equation leads to the result (\ref{Dbfinal}). Equating $\Delta
b_{w}$ determined by the analysis of the stress due to the magnetic
self-force (\ref{DelB}) to (\ref{Dbfinal}) obtained within the theory of
elasticity
\begin{equation}
\frac{wb^{2}}{\pi E_{w}a^{2}}=\frac{\mu _{0}}{4\pi ^{2}}\frac{I^{2}}{a^{2}}%
\frac{b}{E_{w}}\left( \ln \frac{8b}{a}-\frac{3}{4}+2\pi \nu _{w}\right) ,
\end{equation}%
one gets the magnitude of the radial uniformly distributed force $w$ per
units length of the circumference
\begin{equation}
w=\frac{\mu _{0}}{4\pi }\frac{I^{2}}{b}\left( \ln \frac{8b}{a}-\frac{3}{4}%
+2\pi \nu _{w}\right) .  \label{wForce}
\end{equation}%
Equation (\ref{wForce}) concludes the quest for the self-force acting on the
current-carrying wire due to the magnetic field induced by the current. As
could be expected this force per unit of length is proportional to $\frac{%
\mu _{0}}{2\pi }\frac{I^{2}}{(2b)}$ as for two long wires parallel to each
other which are separated by the distance 2$b$ and carrying equal current $I$%
. Every infinitesimally short piece of the circular wire has a counterpart
parallel to it at the diameter 2$b$ away. Both pieces have the same current $%
I$ flow in the opposite directions and repel each other in agreement with
the radially out direction of self-force $w$. Due to the circular shape of
the wire the simple expression $\frac{\mu _{0}}{2\pi }\frac{I^{2}}{(2b)}$ is
modified by the factor in parentheses in Eq. (\ref{wForce}). The first term
of this factor is dominant and is of the order of 10 for $b/a\sim
10^{2}-10^{4}$. The other two terms of this factor are of the order of 1 and
are significantly smaller than $\ln \frac{8b}{a}$. In the equilibrium the
stand-alone circular current-carrying wire under the force $w$ is balanced
by the opposite elastic forces.

\subsection{Stress and strain of circular membrane}

\label{C}

\begin{figure}[t]
\centering
\includegraphics[width=14.0cm]{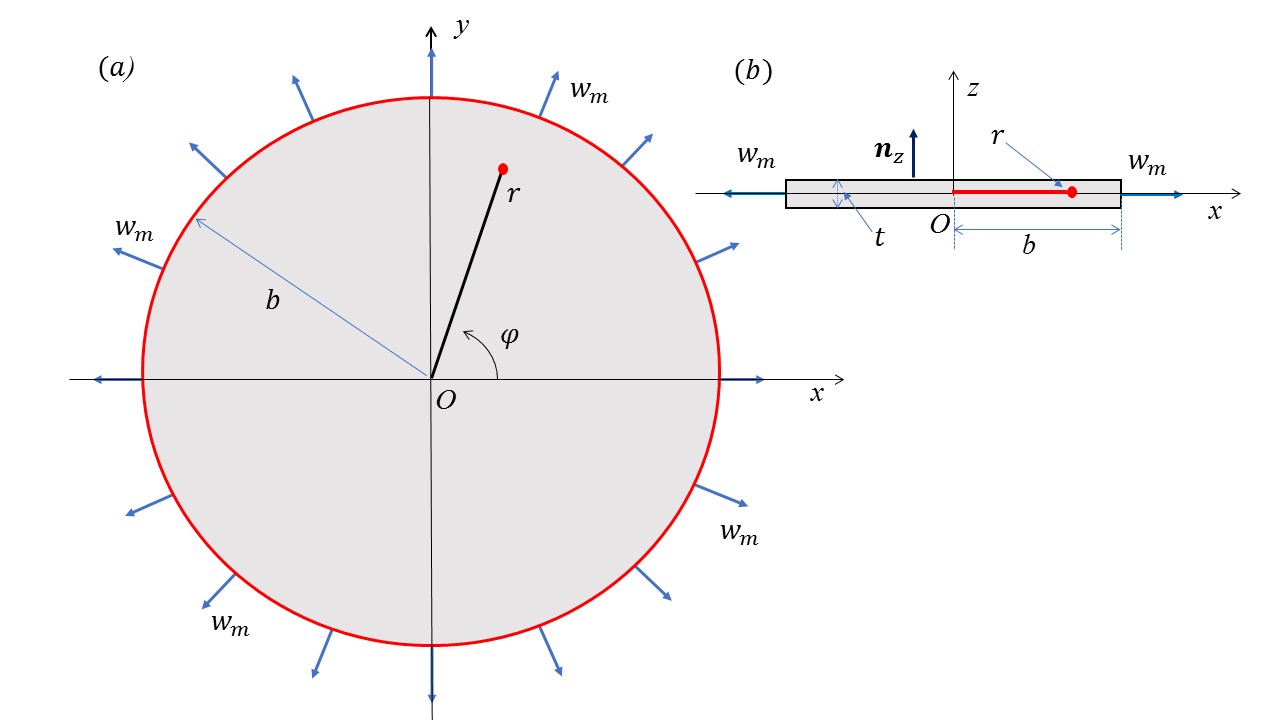}
\caption{(Color online) Schematics for the action of the uniformly
distributed radial force $w_{m}$ per unit of circumferential length of the
circular membrane. ($a$) top view. ($b$) side view. The figure is not to
scale: $b\gg t$.}
\label{Fig6}
\end{figure}

Let's consider a thin circular membrane of the radius $b$ and thickness $t$ (%
$t\ll b$) under a uniform distributed force $w_{m}$ per unit length of a
circumference acting at the membrane edge in the radial direction as shown
in Fig. \ref{Fig6}. If the membrane is sufficiently thin, the deformation
can be treated as uniform over its thickness and we have to deal with
longitudinal deformations of the membrane and not with any membrane bending.
For a two-dimensional case, the strain tensor is a function of $x$ and $y$
coordinates and is independent of $z$. The boundary conditions for the
stress tensor on both surfaces of the membrane are
\begin{equation}
\sigma _{ik}n_{k}=0,n_{k}=n_{z},i=x,y,z,
\end{equation}%
where $n_{k}$ is the normal vector parallel to $z-$axis and lead to
\begin{equation}
\sigma _{xz}=\sigma _{yz}=\sigma _{zz}=0
\end{equation}%
in the whole volume of the membrane when $t\ll b$ \cite{Landau7}.

The equation of equilibrium in the absence of the body forces in the
two-dimensional vector form is \cite{Landau7}:
\begin{equation}
grad \text{ }grad\text{ }\mathbf{u-}\frac{1}{2}(1-\nu _{m}) curl\text{ }curl
\mathbf{u}=0,  \label{Curl}
\end{equation}%
where $\mathbf{u}$ is the displacement vector, $\nu _{m}$ is the membrane
Poisson ratio and all the vector operators are two-dimensional. Due to the
axial symmetry $\mathbf{u}$ is directed along the radius and is a function
of $r$ only, so $curl$ $\mathbf{u}=0$ and Eq. (\ref{Curl}) in polar
coordinates becomes:

\begin{equation}
div\text{ }\mathbf{u=}\frac{1}{r}\frac{d(ru)}{dr}=const\equiv 2c
\end{equation}%
and gives
\begin{equation}
u=cr+\frac{d}{r}  \label{usolutionG}
\end{equation}%
and, therefore, the components are
\begin{equation}
u_{rr}=\frac{du}{dr}=c-\frac{d}{r^{2}},\text{ }u_{\varphi \varphi }=\frac{u}{%
r}=c+\frac{d}{r^{2}}  \label{usolution}
\end{equation}%
where $c$ and $d$ are some constants \cite{Handbook}. In polar coordinates
at $\varphi =0$ the stress $\sigma _{ik}$ and strain $u_{ik}$ tensor
components are:
\begin{eqnarray}
\sigma _{rr} &\mathbf{=}&\sigma _{xx},\text{ }\sigma _{\varphi \varphi }%
\mathbf{=}\sigma _{yy},\text{ }  \label{sigsig} \\
u_{rr} &\mathbf{=}&u_{xx},\text{ }u_{\varphi \varphi }\mathbf{=}u_{yy}.
\label{uu}
\end{eqnarray}%
The general equations relating the strain tensor components to the stress
tensor components \cite{Landau7} in our case of $\sigma _{zz}=0$ become:
\begin{eqnarray}
\sigma _{xx} &=&\frac{E_{m}}{1-\nu _{m}^{2}}\left( u_{xx}+\nu
_{m}u_{yy}\right) ,  \label{Sigxx} \\
\sigma _{yy} &=&\frac{E_{m}}{1-\nu _{m}^{2}}\left( u_{yy}+\nu
_{m}u_{xx}\right) ,  \label{Sigyy}
\end{eqnarray}%
where $E_{m}$ and $\nu _{m}$ are the Young modulus of elasticity and Poisson
ratio of the membrane, respectively. By substituting (\ref{sigsig}) and (\ref%
{uu}) at $\varphi =0$ into (\ref{Sigxx}) and (\ref{Sigyy}) the stress tensor
radial and angular components become:

\begin{eqnarray}
\sigma _{rr} &\mathbf{=}&\frac{E_{m}}{1-\nu _{m}^{2}}\left( u_{rr}+\nu
_{m}u_{\varphi \varphi }\right) ,  \label{sigmarr} \\
\sigma _{\varphi \varphi } &\mathbf{=}&\frac{E_{m}}{1-\nu _{m}^{2}}\left(
u_{\varphi \varphi }+\nu _{m}u_{rr}\right).  \label{sigmarr2}
\end{eqnarray}%
The requirement for the deformation $u$ (\ref{usolutionG}) to be finite at
the membrane center and the boundary condition for $\sigma _{rr}$ (\ref%
{sigmarr}) combined with $u_{rr}$ and $u_{\varphi \varphi }$ (\ref{usolution}%
) at the membrane edge:

\begin{equation}
u\left( 0\right) =0\text{ and }\sigma _{rr}(r=b)=\frac{w_{m}}{t}
\label{bondary}
\end{equation}%
determine the values of the constants $c$ and $d$:%
\begin{equation}
c=\frac{w_{m}}{t}\frac{(1-\nu _{m})}{E_{m}},\text{ }d=0.  \label{const}
\end{equation}%
The use of the constants (\ref{const}) provides the expressions for the
deformation $u$ (\ref{usolutionG}) and the strain (\ref{usolution}) and the
stress (\ref{sigmarr}), (\ref{sigmarr2}) tensor components:%
\begin{eqnarray}
u &\mathbf{=}&\frac{w_{m}}{t}\frac{(1-\nu _{m})}{E_{m}}r, \\
u_{rr} &=&u_{\varphi \varphi }=\frac{w_{m}}{t}\frac{(1-\nu _{m})}{E_{m}},
\label{Strdist} \\
\sigma _{rr} &=&\sigma _{\varphi \varphi }=\frac{w_{m}}{t}.  \label{St45}
\end{eqnarray}%
As expected the stress distribution (\ref{St45}) in the membrane deformed by
the forces acting at the edge of the membrane does not depend on the
elasticity constants of the membrane media \cite{Landau7}. Finally, both the
radial deformation $\Delta b_{m}=u(r=b)$, the strain $\frac{\Delta b_{m}}{b}%
(r=b)$ and the stress $\sigma _{rm}=\sigma _{rr}(r=b)$ of the membrane edge
are:
\begin{eqnarray}
\Delta b_{m} &\mathbf{=}&\frac{w_{m}}{t}\frac{(1-\nu _{m})}{E_{m}}b,
\label{Dbmembrane} \\
\frac{\Delta b_{m}}{b} &=&\frac{w_{m}}{t}\frac{(1-\nu _{m})}{E_{m}},
\label{Dbmembrane1} \\
\sigma _{rm} &=&\frac{w_{m}}{t}.  \label{Dbmembrane2}
\end{eqnarray}%
\begin{figure}[t]
\centering
\includegraphics[width=14.0cm]{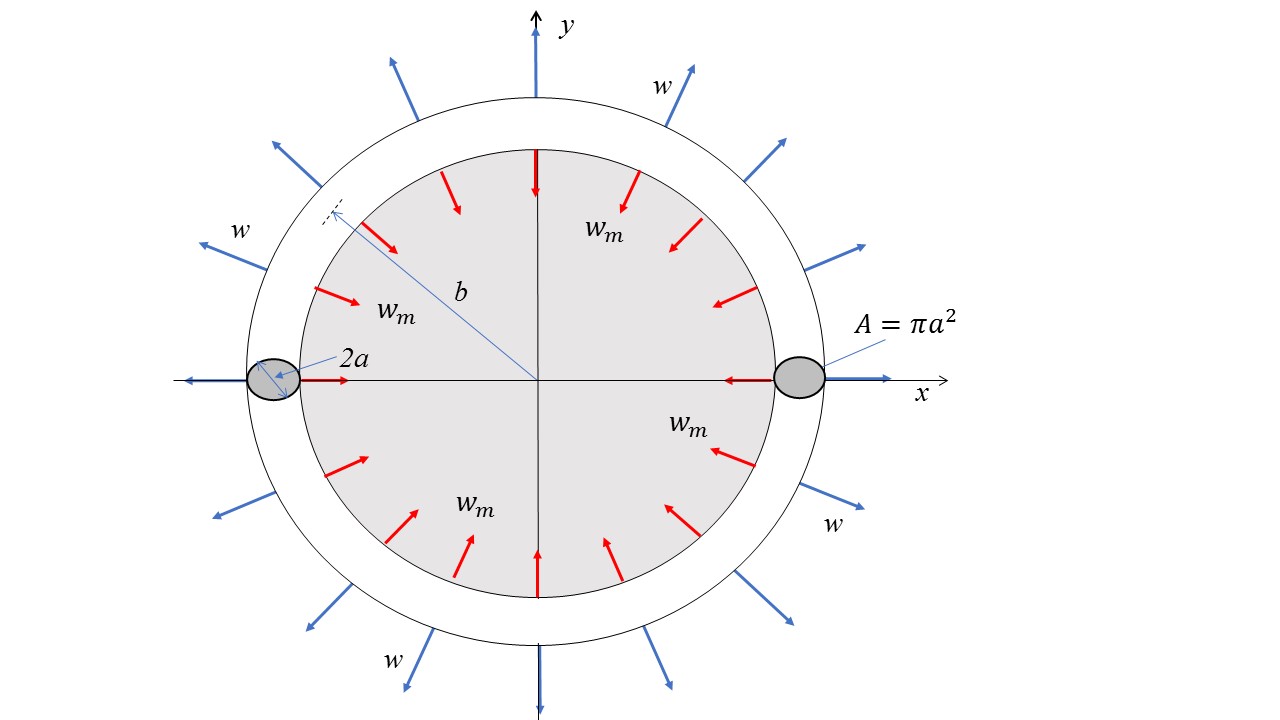}
\caption{(Color online) Schematics for the action of uniform radial forces $%
w $ and $w_{m}$ per unit of circumferential length on the circular
current-carrying wire. The force opposite in direction and equal in
magnitude to $w_{m}$ acts on the circular membrane and is not shown. For
visibility, the cross-section of the wire is given in 3D format. The figure
is not to scale: $b\gg a$.}
\label{Fig5}
\end{figure}
It is important to mention that all expressions in Subsec. \ref{C} are
valid for the membrane of thickness $t$ in the shape of a ring of the
external radius $b$ and internal radius of any value less than $b$ when both
radii have the same center and the membrane is clamped along the internal
edge. This is still the same case of two-dimensional uniform expansion as
considered above.

\subsection{Circular membrane attached to current-carrying wire}

In the absence of the current in the wire, there are no stresses along the
line of the attachment of the membrane to the wire and the membrane forms
the planar disk. The current running in the wire causes the magnetic
self-force $w$ acting on the wire radially out and $w_{m}$ acting on the
wire from the membrane radially in as shown in Fig. \ref{Fig5}. The latter
force is in return to the stress caused by the force of the wire acting on
the membrane and equal in magnitude to $w_{m}$. In the state of equilibrium
we have
\begin{equation}
\Delta b_{w}=\Delta b_{m}.  \label{rav}
\end{equation}%
The radial forces $w$ and $w_{m}$ resulting in total radial force $w - w_{m}$
has to be balanced out by the elastic forces in the wire causing the wire
radius change $\Delta b_{w}$ according to (\ref{Dbfinal}), but with the
force $w$ for stand-alone wire replaced to the force $w-w_{m}$ for the wire
membrane combination. Using (\ref{Dbfinal}) in place of $\Delta b_{w}$ with $%
w$ replaced to $w-w_{m}$, and using (\ref{Dbmembrane}) for the membrane
leads (\ref{rav}) to
\begin{equation}
\frac{(w-w_{m})b^{2}}{\pi E_{w}a^{2}}=\frac{w_{m}}{t}\frac{(1-\nu _{m})}{%
E_{m}}b,
\end{equation}%
and

\begin{equation}
w_{m}=w\left( 1+\frac{\pi a^{2}}{tb}\frac{E_{w}}{E_{m}}(1-\nu _{m})\right)
^{-1}.  \label{wm}
\end{equation}%
In conclusion, (\ref{Dbmembrane1}) with (\ref{wm}) for $w_{m}$ and (\ref%
{wForce}) for $w$ allow the determination of the strain of both the wire and
the membrane, $\sigma _{rm}=w_{m}/t$ (\ref{Dbmembrane2}) determines the
radial stress on the membrane, and (\ref{26}) with $w$ replaced by $w-w_{m}$%
:
\begin{equation}
\sigma _{t}=\frac{\left( w-w_{m}\right) b}{\pi a^{2}}  \label{sigmat}
\end{equation}%
provides for the tensile stress of the wire.

\section{Results and discussion}

\label{results} The theory given in the previous section provides
expressions to calculate different quantities related to solar sail with
superconducting circular current-carrying wire such as self-inductance,
magnetic energy, stress and strain. The general assumptions of applicability
of these expressions are the following:

i. the materials of the wire and the membrane are nonpermeable;

ii. the radius of wire cross-section $a\ll b$, where $b$ is the radius of
both the wire circular loop and membrane;

iii. both strain and stress are small and within the limit of elasticity.

The first condition is satisfied when magnetically weak materials are used,
such as a superconducting wire. To satisfy the second condition the values
of $\frac{a}{b}\leq 2\times 10^{-3}$ will be used. The third condition means
that the stress in the wire and membrane have to be less than the yield
strength of the corresponding material.

Let's consider CP1 Polyimide films \cite{CP1} as an example of the material
of the membrane used for the solar sail. The properties of CP1 are listed in
Table \ref{tab1}. The yield strength of CP1 is not available to the authors.
In general, it is much less than the Young modulus. Not knowing better one
can estimate the yield strength to be greater than 10$^{-3}$ part of the
Young modulus, keeping in mind that the real design decisions have to be
based on experimentally measured values.

\begin{table}[h]
\caption{Model parameters for the superconducting wire extracted from Refs.
\protect\cite{45,46}, and the sail membrane extracted from Refs.
\protect\cite{CP1}. $\protect\rho _{w}$ and $\protect\rho _{m}$, $E_{w}$ and
$E_{m}$, $\protect\nu _{w}$ and $\protect\nu _{m}$ are the density, Young
modulus, Poisson ratio, for the superconducting wire and sail membrane,
respectively. $t$ is the membrane thickness. }
\label{tab1}
\begin{center}
\begin{tabular}{ccc|cccc}
\hline\hline
\multicolumn{3}{c|}{Superconducting wire, Bi$-$2212} & \multicolumn{4}{|c}{
Sail's membrane, CP1} \\ \hline
$\rho _{w}$, kg/m$^{3}$ & $E_{w},$ Pa & $\nu _{w}$ & $\rho _{m}$, kg/m$^{3}$
& $E_{m}$, Pa & $\nu _{m}$ & $t$, m \\ \hline
$9.5\times 10^{3}$ & $8.0\times 10^{10}$ & $3.7\times 10^{-1}$ & $1.43\times
10^{3}$ & $2.17\times 10^{9}$ & $3.40\times 10^{-1}$ & $3.5\times 10^{-6}$
\\ \hline
\end{tabular}%
\end{center}
\end{table}

The High Temperature Superconducting (HTS) wire made of (Bi,Pb)$_{2}$Sr$_{2}$%
Ca$_{3}$Cu$_{2}$O$_{8-x}$ (Bi$-$2212) with properties listed in Table \ref%
{tab1} is chosen as the current-carrying wire. The Bi$-$2212 wire can be
made into round-wire, multifilamentary strand embedded in strengthen Ag
matrix \cite{CSWire,45}. According to the yield strength data presented in
Ref. \cite{45} it is also safe to
estimate the yield strength for Bi$-$2212 superconducting wire to be greater
than 10$^{-3}$ part of the corresponding Young modulus. The estimate of the
engineering current density $J_{e}$, which is the current density over the
whole wire cross-section including Ag matrix, is in the range of 500 A/mm$%
^{2}$ to 1000 A/mm$^{2}$.
The rapid progress in the development of HTS wires allows to be optimistic
about the commercial availability of HTS wires with even greater $J_{e}$ in
near future. The current in the wire $I=\pi a^{2}J_{e}$ will be used in the
calculations.

\begin{figure}[b]
\centering
\includegraphics[width=6.0cm]{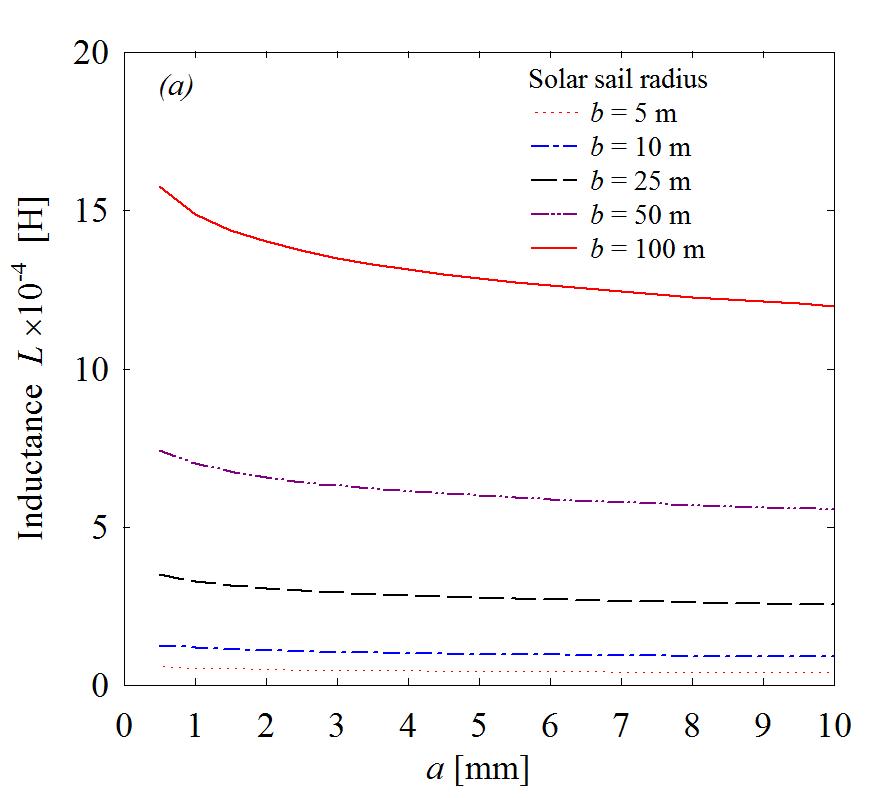} \includegraphics[width=6.3cm]{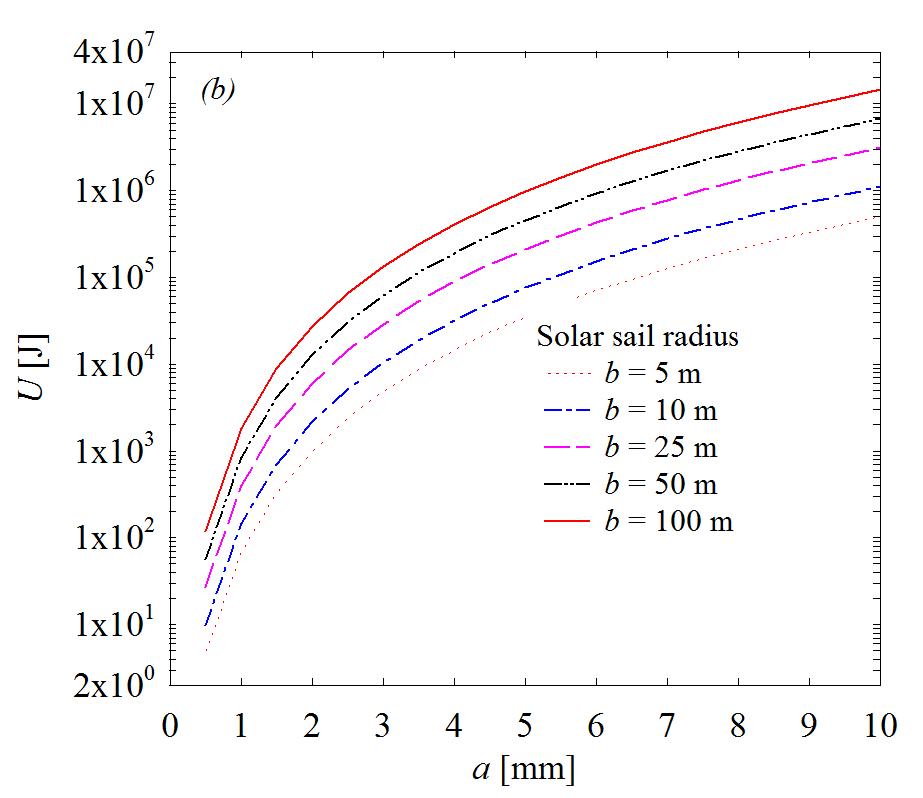}
\caption{(Color online) Dependence of the superconducting wire
self-inductance $L$ ($a$) and the total energy $U$ of the magnetic field of
the wire ($b$) on the radius of the wire. Calculations performed for the
engineering current $J_{e} = 500$ A/mm$^{2}$ and a solar sail with areas
from 78.5 m$^{2}$ to 10$^{4}$ m$^{2}$.}
\label{FigV7}
\end{figure}

\begin{figure}[h]
\centering
\includegraphics[width=7.8cm]{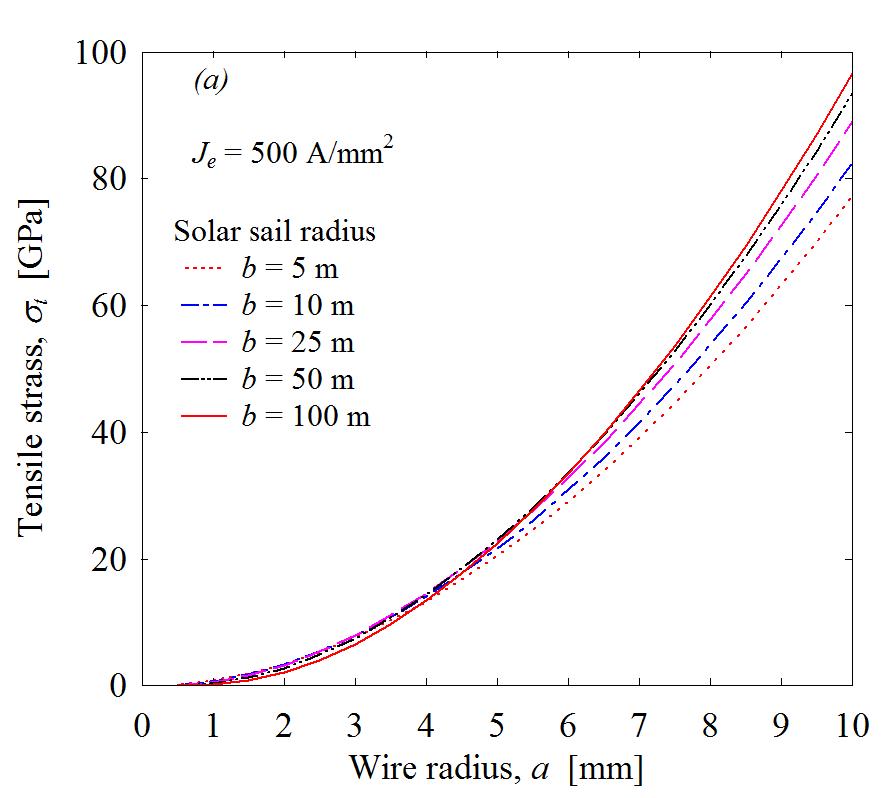} \includegraphics[width=7.8cm]{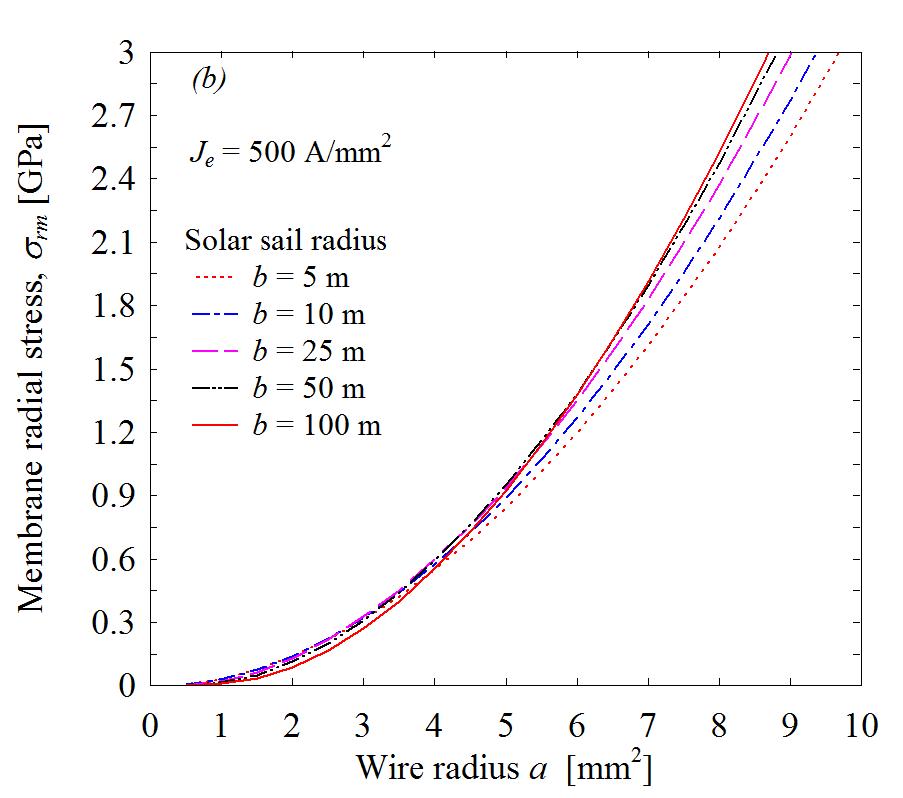} %
\includegraphics[width=7.9cm]{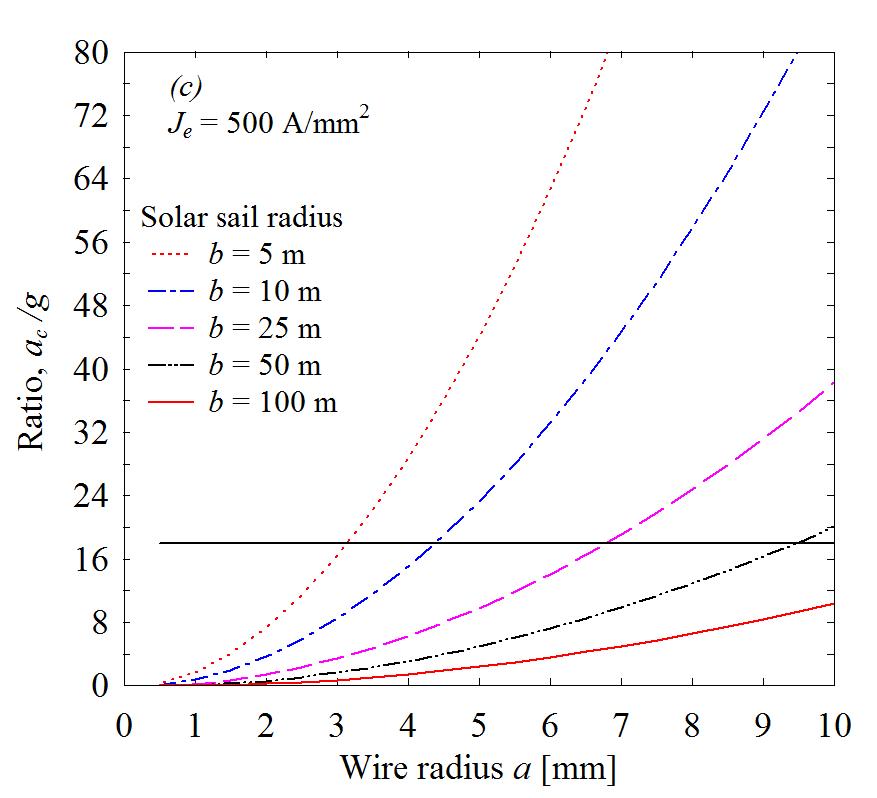} \includegraphics[width=7.8cm]{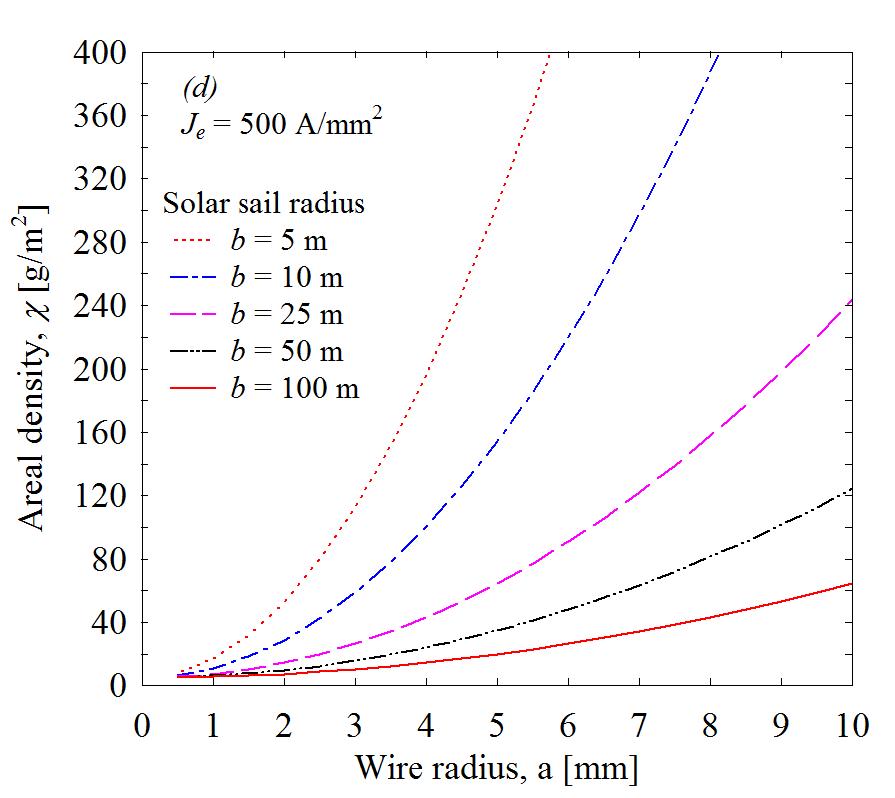}
\caption{(Color online) Dependence of the tensile stress ($a$), membrane
radial stress ($b$), the ratio $a_{c}/g$ ($c$), and arial density ($d$) on
the wire radius. Calculations performed for the current density $J_{e}=500$
A/mm$^{2}$ and different sizes of solar sail. The horizontal line in ($c$)
for the ratio $a_{c}/g$ is obtained based on the result from Ref.
\protect\cite{Deployment4}.}
\label{FigV8}
\end{figure}
The expressions (\ref{LSC}) and (\ref{Total E}) are used to calculate the
superconducting wire self-inductance $L$ and the total energy $U$ of the
magnetic field of the wire carrying current $I$, respectively. The twofold
of that energy $2U$ determines the minimum requirement of energy consumption
of the source of \textit{EMF} to establish current $I$ in the wire. The
expression (\ref{wForce}) is employed to calculate the self-force $w$ acting
on the current-carrying wire due to the magnetic field induced by the
current, and expression (\ref{wm}) gives the value of the force $w_{m}$
acting at the membrane edge. The knowledge of $w$ and $w_{m}$ is used to
calculate the stress of the membrane $\sigma _{rm}$ using Eq. (\ref%
{Dbmembrane2}), the stress of the wire $\sigma _{t}$ using Eq. (\ref{sigmat}%
) and the strain $\Delta b_{m}/b$ of both Eq. (\ref{Dbmembrane1}).

Let us also introduce the ratio $a_{c}/g$ given as

\begin{equation}
\frac{a_{c}}{g}=\frac{\left( w-w_{m}\right) }{\rho _{w}\pi a^{2}g},
\label{ratio}
\end{equation}%
where $\rho _{w}$ is the wire mass density and $g=9.8$ m/s$^{2}$ is the
acceleration due to gravity at the Earth surface. The quantity $a_{c}$ is
effective acceleration equivalent to centripetal acceleration of the
circular wire would it be spinning around its center with frequency
\begin{equation}
f=\frac{1}{2\pi }\left( \frac{a_{c}}{g}\right) ^{1/2}.
\end{equation}%
The greater $a_{c}/g$ ratio the greater the chance of successful deployment
of initially folded superconductive wire and sail membrane to the open state
of circular shape. A simple analysis of the partial derivatives shows that $%
\sigma _{t}$, $\sigma _{rm}$, and $a_{c}/g$ are the monotonically increasing
functions of radius $a$ of the wire cross-section and current density $J_{e}$%
. The behavior of $\sigma _{t}$ and $\sigma _{rm}$, when the radius $b$ of
the wire varies is not that simple. Both $\sigma _{t}$ and $\sigma _{rm}$
exhibit a broad maximum at the same value of $b$. The position of the
maximum shifts to the greater value of $b$ with an increasing value of $a$.
The $a_{c}/g$ ratio behavior remains simple, it decreases monotonically with
an increase of $b$.

One of the key metrics related to the performance of solar sail is the
characteristic acceleration \cite{Colin,Matloff,Matloff2}. The characteristic acceleration is
defined as $a_{0}=2\eta P_{0}/\chi $, where $0.5\leq \eta \leq 1$, $P_{0}$
is the solar radiation pressure near the Earth and $\chi =m/A$ is the areal
density. In the latter expression, $m$\ is the total mass of the solar sail,
in our case a sum of the wire and sail membrane masses, and $A$ is the sail
membrane area. While the actual sail acceleration is a function of
heliocentric distance and its orientation, the characteristic acceleration
allows a comparison of solar sail design concepts on an equal footing. The
areal density $\chi $ is an important parameter determining the sail
acceleration due to light pressure. It is easy the calculate $\chi $ for
known geometry and densities of sail materials. Less $\chi $ leads to the
greater sail acceleration.

\begin{figure}[b]
\centering
\includegraphics[width=7.8cm]{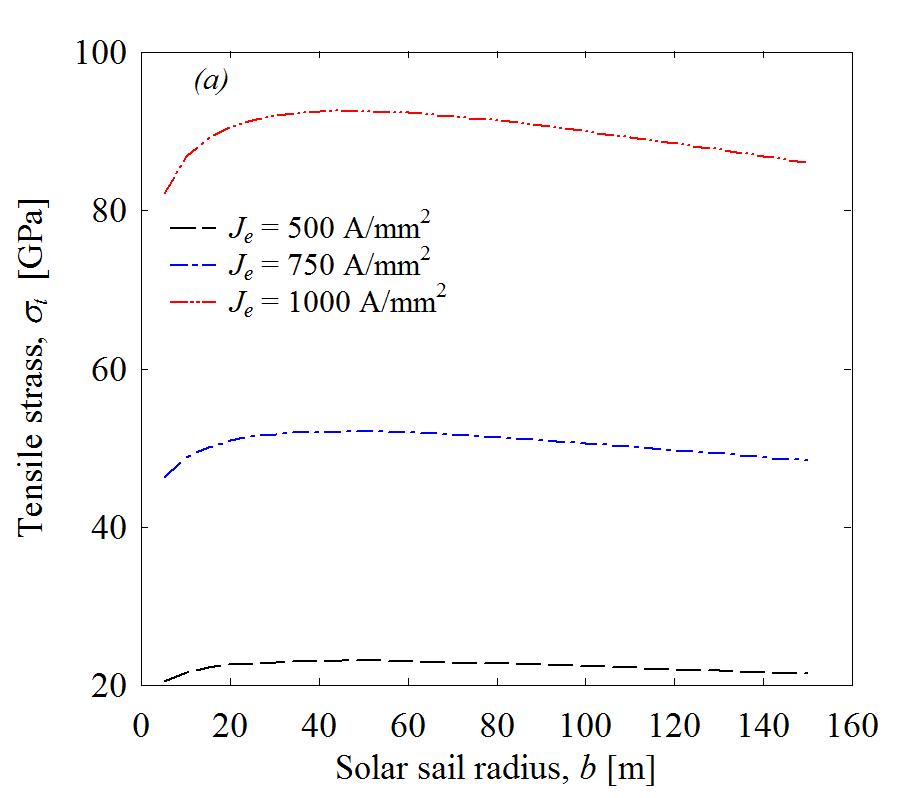} \includegraphics[width=7.8cm]{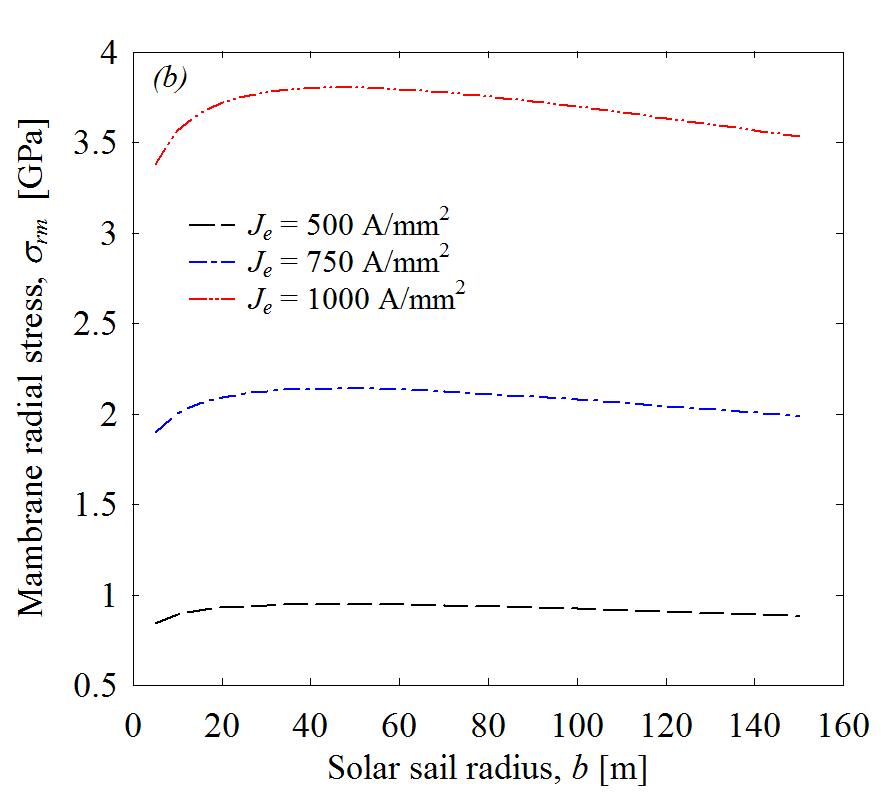} %
\includegraphics[width=7.8cm]{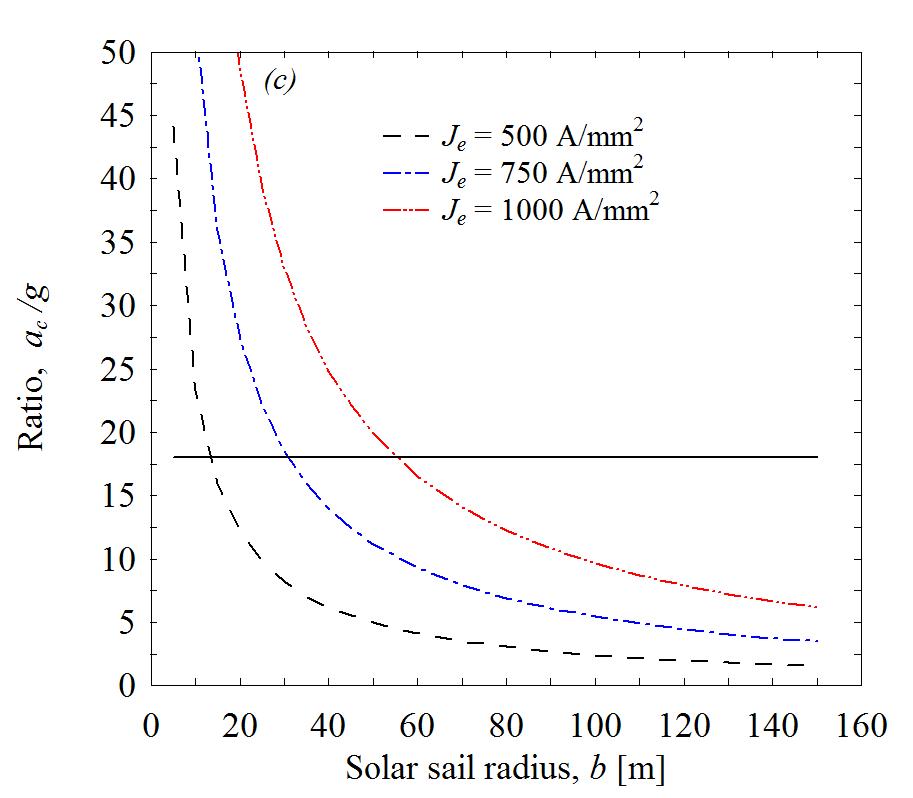} \includegraphics[width=7.8cm]{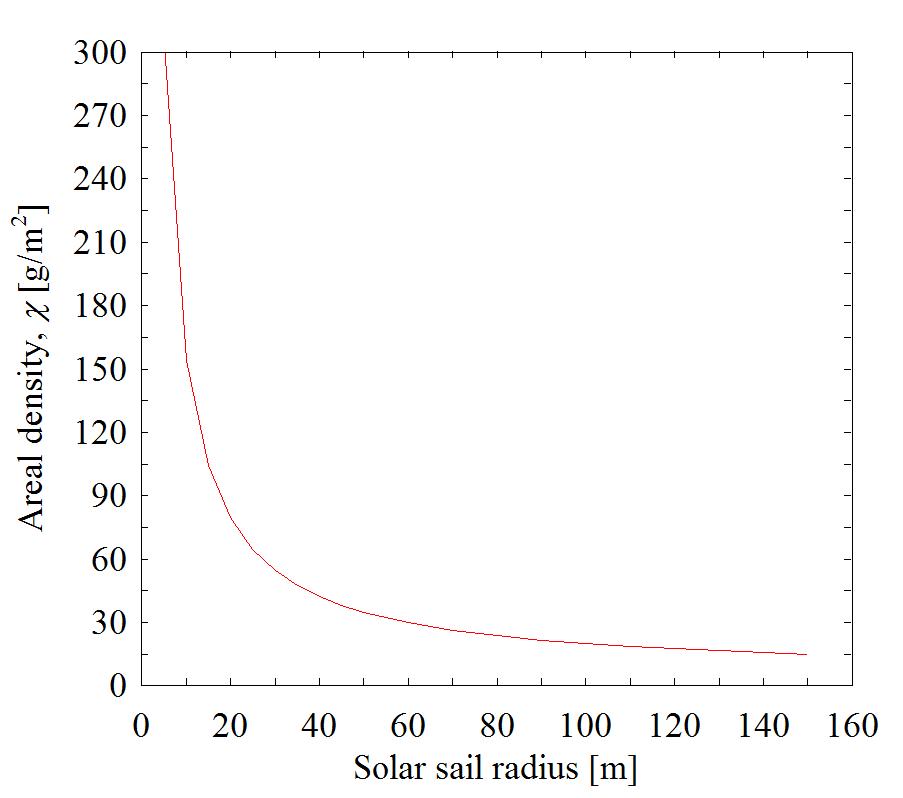}
\caption{(Color online) Dependence of the tensile stress ($a$), membrane
radial stress ($b$), the ratio $a_{c}/g$ ($c$), and arial density ($d$) on
the solar sail radius. Calculations performed for the wire radius $a=5$ mm
and different current densities $J_{e}$. The horizontal line in ($c$) for
the ratio $a_{c}/g$ is obtained based on the result from Ref. \protect\cite%
{Deployment4}.}
\label{FigV9}
\end{figure}

The self-inductance $L$ and magnetic energy $U$ are presented in Fig. \ref%
{FigV7} as the function of the radius $a$ for several values of $b$. The
values of $L$ are dependent only on the geometry. The values of $U$ are
given for the engineering current density $J_{e}=500$ A/mm$^{2}$ and being
dependent on the square of the current can be used to find the energy for
another current density. The energy 2$U$ is of special interest giving the
minimal requirement for the power supply. The superconducting wire can be
operated in persistent mode with the power supply turn off after the current
is established in the wire. The minimal estimate for power supply is as high
as 2 MJ for the $b=10$ m sail radius with $a=10$ mm wire radius.

The dependencies of stresses $\sigma _{t}$, $\sigma _{rm}$, $a_{c}/g$ ratio,
and the areal density $\chi $, respectively, on the wire radius $a$ for $%
J_{e}=500$ A/mm$^{2}$ and different fixed values of solar sail radius are
presented in Fig. \ref{FigV8}. It is clear that both the wire and membrane
stresses stay within the 10$^{-3}$ range of the corresponding Young modulus
for $a<7$ mm. Thus $a=7$ mm is the maximum radius of the wire cross-section
to be safely used for the choice of $b<100$ m and $J_{e}=500$ A/mm$^{2}$.
The $a_{c}/g$ ratio values range depends greatly on the value of the sail
radius $b$. The greater the value of $b$ the less the range. For example,
for $b=100$ m the ratio stays below 10. For $b=$ 5 m the ratio increases
from 0.3 up to 63 as the wire radius $a$\ changes from 0.5 mm to 6 mm.

The logical question which arises now is the following: which value of the
ratio is enough to deploy the sail? Although a certain answer can be
obtained only in the experiment it's possible to compare the calculated
values of $a_{c}/g$ ratio with the equivalent estimate based on the
successful ground demonstration of the spinning sail deployment by Jet
Propulsion Laboratory \cite{Deployment4}. The data provided in Ref. \cite%
{Deployment4} for the spinning frequency 200 RPM or more
of the sail of 0.8 m in diameter gives the estimate of the ratio of
centripetal acceleration over the acceleration due to gravity equal to 18 or
more. The calculated $a_{c}/g$ ratio presented in Fig. \ref{FigV8}c exceeds
this estimate for the certain range of wire radius $a$ values for a choice
of sail radius $b<$ 25 m. For example, the ratio is well above the estimate
for $b=10$ m as $a>4$ mm. As is expected the areal density increases with
the increase of the wire radius and decrease of the sail radius. This means
that for the certain choice of sail size the less value of wire radius is
preferable as far as the $a_{c}/g$ ratio allows the sail to be deployed and
the wire and the sail membrane stay within the limit of elasticity. For $%
J_{e}=$500 A/mm$^{2}$ and $b=10$ m the choice of $a=5$ mm is close to an
optimal with $\chi =155$ g/m$^{2}$.

Figure \ref{FigV9} presents the stresses $\sigma _{t}$, $\sigma _{rm}$, $%
a_{c}/g$ ratio, and the areal density $\chi $, respectively, as function of
sail radius $b$ for the different current densities $J_{e}$ and the fixed
value of the wire radius $a=5$ mm. It is clear that both the wire and
membrane stresses stay within the 10$^{-3}$ range of the corresponding Young
modulus for $J_{e}\leq 750$ A/mm$^{2}$. As expected, both the $a_{c}/g$ and $%
\chi $ monotonically decrease with the increase of the sail radius $b$. For
fixed $b$ the increase of current density leads to increase of the $a_{c}/g$
without any effect of the areal density. This means that the reasonable
values of $b$ increase with the increase of the current density. For
example, for the chosen values of $a=5$ mm and $J_{e}=750$ A/mm$^{2}$ the
solar sail radius can be as big as 25 m with $\chi \sim 64$ g/m$^{2}$.

It is worth noting that we performed the same calculations for
the dependencies of stresses on the wire and solar sail radii
for the most widely used low temperature superconductor Nb$-$Ti. The similar results were obtained.

\section{Concluding remarks}

\label{conclusions}

The theory of circular solar sail attached to superconducting
current-carrying wire is developed within the framework of classical
electrodynamics and theory of elasticity. We obtained the analytical
expressions that can be applied to a wide range of the materials of both the
wire and sail membrane. The presented numerical example demonstrates the
power of the developed theory to provide sound estimates of the stresses of
the solar sail with attached superconducting circular current-carrying wire
along with the power requirements solely based on the geometry, engineering
current density and elastic properties of the materials used.

It is possible to imagine different options of the sail attachment to the
useful load, which can represent the spacecraft itself. The theory presented
is directly applicable to the case when the sail is in the shape of a ring
and is clamped along the internal edge to the load having a circular cross
section in the plane of the sail. Another option is the sail in the shape of
the rosette of the daisy. In this case, the leaves of the rosette are sails
of the same size arranged regularly around the base represented by a load of
circular cross section. Each leave sail is the circular sail with
superconducting current-carrying wire around. It is not necessary to have
all the leave sails in the same plane, each of them can have its own
orientation in space allowing to steer the load. Moreover, one can envision
that the orientation of each leave can be changed to achieve the needed
maneuver. Arrangement of multiple sails in daisy-like shape will require
further advance of the presented theory to take into account the effects of
mutual inductance of the sails. The case of the load attached via flexible
tethers to the wire of the sail at regular length intervals will require the
theory advance to take into account the action of the forces at the points
of tethers attachment. One can operate the sail in persistent mode when the
superconducting wire is detached from the power supply after the current is
established, and also dump the current and close the sail using the
lightweight flexible tethers and re-open it by injecting the current in the
wire as needed. This kind of close-open operating mode can be quite useful
in the case of daisy shaped sail, when all the leave sails are in the same
plane, but the needed maneuver can be achieved by closing one or more leave
sails and later re-opening them.

Today all launched solar sails have a square shape and such design is related to the deployment mechanism. The world's first interplanetary solar sail, -- the IKAROS, successfully deployed its 196 m$^{2}$ sail in 2010. NASA's first solar sail deployed in low earth orbit was NanoSail-D which had 9.3 m$^{2}$ of light-reflecting catching surface and the LightSail-2 on July 23, 2019, deployed its 32 m$^{2}$ solar sail. For comparison, our calculations show that the sail of $\sim 1,963$ m$^{2}$ area (25 m radius) with the attached wire of 5 mm cross-section radius is expected to be deployed by the current in the wire of the engineering density 750 A/mm$^{2}$ with $a_{c}/g = 22$ and $\chi=64$ g/m$^{2}$. This sail is even bigger than NASA's Solar Cruiser, a mission launching in 2025 to test a giant sail measuring 1,650 m$^{2}$.

To conclude our calculations demonstrate the feasibility of the proposed
idea for the deployment and stretching of the circular solar sail
constructed as the superconducting current loop attached to the thin
circular membrane. At this point, the sound experimental study is needed to
find out the level of the feasibility of the proposed idea for the future of
the deep space exploration by means of the light pressure propellent.

\acknowledgments

We are grateful to Bernd Dachwald (FH Aachen University of Applied
Sciences), Les Johnson (NASA, George C. Marshall Space Flight Center), and
Greg Matloff (City Tech, CUNY) for providing useful information.

\end{document}